\documentstyle[preprint,aps,epsfig]{revtex}
\begin{document}

\tightenlines

\preprint{\vbox{\hbox{BNL-64628}
\hbox{PRINCETON/HEP/97-10}
\hbox{TRI--PP--97--26}
\hbox{KEK Preprint 97-113}}}

\draft

\title{Observation of the Decay $K^+\rightarrow\pi^+\gamma\gamma$}

\author{P. Kitching, T. Nakano\cite{TN}, M. Rozon\cite{MR}, and
  R. Soluk}
\address{
  Centre for Subatomic Research, University of Alberta, Edmonton, Alberta, Canada, T6G 2N5
}

\author{S. Adler, M.S. Atiya, I-H. Chiang, J.S. Frank, J.S. Haggerty,
  T.F. Kycia, K.K. Li, \\ L.S. Littenberg, A. Sambamurti\cite{AKS},
  A. Stevens, R.C. Strand, and C. Witzig}
\address{
  Brookhaven National Laboratory, Upton, New York 11973
}

\author{W.C. Louis}
\address{
  Medium Energy Physics Division, Los Alamos National Laboratory, \\
Los Alamos, New Mexico 87545
}

\author{D.S. Akerib\cite{DSA}, M. Ardebili\cite{MA}, M. Convery\cite{MC}, 
M.M. Ito\cite{ITO}, D.R.~Marlow, R.A.~McPherson\cite{ROB}, 
P.D.~Meyers, M.A. Selen\cite{MAS}, F.C. Shoemaker, and A.J.S. Smith}
\address{
  Joseph Henry Laboratories, Princeton University,
Princeton, New Jersey 08544
}

\author{E.W. Blackmore, D.A. Bryman, L. Felawka, 
  A. Konaka, Y. Kuno\cite{YK}, J.A. Macdonald, \\ T. Numao,
  P. Padley\cite{PP}, J.-M. Poutissou, R. Poutissou, J. Roy\cite{JR},
  and A.S. Turcot\cite{AST}}
\address{
  TRIUMF, Vancouver, British Columbia, Canada, V6T 2A3
}


\date{August, 1997}
\maketitle

\begin{abstract}
The first observation of the decay $K^+\rightarrow\pi^+\gamma\gamma$ is
reported.  A total of 31 events was observed with an estimated
background of 5.1$\pm$3.3 events in the $\pi^{+}$ momentum range from
100 MeV/$c$ to 180 MeV/$c$. The corresponding partial branching ratio,
B($K^+\rightarrow\pi^+\gamma\gamma$, 100 MeV/$c$ $ < P_{\pi^+} < $ 180
MeV/$c$), is $(6.0\pm1.5 \mbox{\{stat\}} \pm0.7 \mbox{\{sys\}})\times
10^{-7}$. No $K^+\rightarrow\pi^+\gamma\gamma$ decay was observed in
the $\pi^+$ momentum region greater than 215 MeV/$c$.  The observed
$\pi^+$ momentum spectrum is compared with the predictions of chiral
perturbation theory.
\end{abstract}



We report the first observation of the rare decay
$K^+\rightarrow\pi^+\gamma\gamma$. This decay provides a stringent
test of chiral perturbation theory (ChPT) \cite{cpta} since there is
no tree-level $O(p^2)$ contribution and the leading contributions
start at $O(p^4)$ \cite{ecker}. It includes one undetermined
coupling constant, $\hat{c}$, at $O(p^4)$ for which various models 
predict different values of order of 1
($O(1)$)\cite{wdm,lnc,ntlo,fact}. Both the branching ratio and the
spectrum shape of $K^+\rightarrow\pi^+\gamma\gamma$ are sensitive to
$\hat{c}$, and the total branching ratio is predicted to be $(0.4 -
1)\times 10^{-6}$ for $\hat{c}$ of $O(1)$.  Measurements of
$K_L\rightarrow\pi^0\gamma\gamma$ \cite{na31,e731} indicate the
necessity for next-to-leading order ($O(p^6)$) ``unitarity'' corrections
\cite{kla,klb} which can be deduced from an empirical fit of the decay
amplitude of $K_L \rightarrow \pi^+ \pi^-
\pi^0$.  Similar corrections for
$K^+\rightarrow\pi^+\gamma\gamma$ from $K^+
\rightarrow \pi^+ \pi^- \pi^+$ predict a 30--40\% higher branching
ratio than the $O(p^4)$ calculations with $\hat{c} \sim O(1)$ \cite{uni}. 
From the measurement of the branching ratio and spectrum shape of
$K^+\rightarrow\pi^+\gamma\gamma$, it is possible to determine a value
of $\hat{c}$ and also to examine whether the unitarity corrections are
necessary.  Observation of this decay mode in excess of the Standard
Model rate would suggest the existence of new phenomena such as
sequential decays of the form $K^+ \rightarrow \pi^+ X^0,X^0
\rightarrow \gamma\gamma$, where $X^0$ is a massive short-lived
neutral scalar particle. Our previous search\cite{ypigg} for
$K^+\rightarrow\pi^+\gamma\gamma$ was confined to the region of phase
space with $P_{\pi^{+}} > 215$ MeV/$c$. We obtained a 90\%
C.L. upper limit on B($K^+\rightarrow\pi^+\gamma\gamma$) of $10^{-6}$
when a constant matrix element (phase-space distribution) was assumed,
but only $10^{-3}$--$10^{-4}$ when the ChPT spectrum with $\hat{c} \sim
 O(1)$ was assumed. The present work expands the search to include
the region $P_{\pi^{+}}$ $<$ 180 MeV/$c$ which is favored by ChPT.

The E787 setup at the Brookhaven Alternating Gradient Synchrotron has been
described in detail in Ref.\cite{e787nim}. An 800-MeV/$c$ kaon beam
was slowed by a BeO degrader and stopped in a segmented scintillating
fiber target located at the center of the detector.  Charged decay
products were momentum-analyzed in a cylindrical drift chamber placed
in a 1-T solenoidal magnetic field.  Their kinetic energy and range
were measured by using the target and a 15-layer plastic scintillator
(range stack) which surrounded the drift chamber.  Signals from the
range stack were fed to 500 MHz transient digitizers (TD) in order to
identify the $\pi^+
\rightarrow \mu^+ \rightarrow e^+$ decay chain in the stopping layer.  
Photons were detected
in a lead-scintillator barrel calorimeter surrounding the range stack.
The barrel calorimeter had 48-fold azimuthal and 4-fold radial
segmentation with a total thickness of 14 radiation lengths. It
covered a solid angle of about $3 \pi$ sr.  Most of the remaining 
solid angle in the end regions was covered by
12-radiation-length end-cap calorimeters located upstream and 
downstream of the drift chamber.

To avoid background from the two-body decay $K^+\rightarrow\pi^+\pi^0$
($K_{\pi 2}$) with a 21\% branching ratio and a 
monochromatic $\pi^+$ momentum of 205 MeV/$c$
($K_{\pi 2}$ peak), we searched for the
$K^+\rightarrow\pi^+\gamma\gamma$ decay in the regions above the
$K_{\pi 2}$ peak, $P_{\pi^{+}} > 215$ MeV/$c$ ($\pi\gamma\gamma1$
region), and below the peak, 100 MeV/$c$ $<$ $P_{\pi^{+}}$ $<$ 180
MeV/$c$ ($\pi\gamma\gamma2$ region).  To select
$K^+\rightarrow\pi^+\gamma\gamma$ candidates, multilevel triggers
followed a threefold strategy: (1) a $K^+$ must stop in the target;
(2) the decay particle must penetrate into the range stack with the
range longer ($\pi\gamma\gamma1$) or shorter ($\pi\gamma\gamma2$)
than that of a $\pi^+$ from $K_{\pi 2}$, and be identified as a $\pi^+$; 
and (3) the number of photon
clusters in the barrel calorimeter must be exactly two with two-photon
invariant mass ($m_{\gamma\gamma}$) smaller ($\pi\gamma\gamma1$) or
larger ($\pi\gamma\gamma2$) than the $\pi^0$ mass.  The
$\pi\gamma\gamma1$ trigger which satisfied the above
strategy was achieved by modifying the logic described in Ref.  
\cite{ypigg} to reject events with photons showering in
the end-cap calorimeters and events with $m_{\gamma\gamma}$ greater than 130
MeV/$c^2$ ($P_{\pi^{+}}=207$ MeV/$c$).  
For the search in the $\pi\gamma\gamma2$ region,
requirements in addition to the beam and range conditions at the first
trigger level were (i) a large energy deposit ($\sim 190$ MeV)
in the barrel calorimeter and (ii) the large $\gamma$--$\gamma$
opening angle ($\theta_{\gamma\gamma} > 75 \deg$) so as to reduce
$K_{\pi 2}$ background. At the second trigger level, the on-line TD
requirement to identify the $\pi^+ \rightarrow \mu^+ \nu_{\mu}$ decay
\cite{ypigg} was removed to avoid an unnecessary acceptance loss.  At
the third trigger level, $m_{\gamma\gamma}$ was calculated, and an
event was rejected if $m_{\gamma\gamma} < 190$ MeV/$c^2$ 
($P_{\pi^{+}}=183$ MeV/$c$).  
From the data taken in 1991, the total number of stopped kaons was
$3.1 \times 10^{10}$ for the $\pi\gamma\gamma1$ trigger and $6.1
\times 10^{10}$ for the $\pi\gamma\gamma2$ trigger.  The difference
between the two $K^+$ counts was mainly due to a hardware prescale
factor of 2 applied to the $\pi\gamma\gamma1$ trigger. A total of $7.3
\times 10^{5}$ events survived the $\pi\gamma\gamma1$ trigger, and
$2.7 \times 10^{6}$ events survived the $\pi\gamma\gamma2$ trigger.

For the off-line search in the $\pi\gamma\gamma1$ region \cite{adler},
we required unambiguous identification of a single charged track as a
$\pi^+$ by measuring its range, energy and momentum, and by observing
the $\pi^+
\rightarrow \mu^+ $ decay. Adjacent hit modules in the barrel
calorimeter were grouped to identify two isolated photon showers
(clusters). The hit position in each module along the beam axis ($z$)
was calculated from the end-to-end time and energy differences. The
azimuthal angle ($\phi$) of the hit position was determined up to the
segmentation of the modules.  Then, the location of the photon shower
in $z$ and $\phi$ was obtained by an energy-weighted average of the
hit positions.  Events were rejected if the photon-pair opening angle
and an energy of either the high or low energy photon were consistent with
$K^+\rightarrow\pi^+\pi^0$, $\pi^0\rightarrow\gamma\gamma$
kinematics. Cuts were applied on the photon energy ratio and opening 
angle to remove events which could be affected by fluctuations in visible 
energy or by accidentals.
  Finally, the remaining events were tested by a constrained
kinematic fit for consistency with
$K^+\rightarrow\pi^+\gamma\gamma$. The constraints included total
momentum and energy conservation and consistency of the charged track's
energy, momentum and range with a pion hypothesis. Events that had a
$\chi^2$ confidence level, P($\chi^2$), greater than 0.1 were
accepted. No event was seen in the region of $E_{\pi^+} > 117$
MeV/$c^2$, or equivalently $P_{\pi^+} > 215$ MeV/$c$. 

The analysis for the $\pi\gamma\gamma2$ region followed procedures
mostly similar to those of the $\pi\gamma\gamma1$ analysis with the
following differences. The major background
sources in the $\pi\gamma\gamma2$ region are the $K^+ \rightarrow
\pi^+ \pi^0 \pi^0$ ($K_{\pi 3}$) and $K^+ \rightarrow \pi^+ \pi^0
\gamma$ ($K_{\pi 2\gamma}$) decays with escaping or overlapping photons;
these channels do not involve muons in the decay products. 
This allowed removal of the $\pi^+ \rightarrow \mu^+$ tagging condition
in order to increase the acceptance by more than a factor of 2. Since there was
a large correlation between the measured energy and range for a low energy
$\pi^+$, the range-momentum relation was removed from the constraints
of the kinematic fit.  Photon vetoes were tightened to reject $K_{\pi
3}$ and $K_{\pi 2\gamma}$ events with escaping photons.  A cut was
placed on the angle between the $\pi^+$ and the higher-energy photon
($\theta_{\pi^+\gamma_1}$) to remove events with a highly asymmetric
photon-pair decay.  Events were rejected if $\cos
\theta_{\pi^+\gamma_1} < -0.8$.  This cut also reduced $K_{\pi
2\gamma}$ events with the radiative photon overlaying one of the two
photons from $\pi^{0}$ decay. Since the radiative photon from $K_{\pi
2\gamma}$ was preferentially aligned along the the $\pi^+$ direction, an
overlapping-photon event usually has one $\pi^0$ photon near the
$\pi^+$ direction and the other in the opposite direction. Events were 
accepted if the constrained kinematic fit satisfied P$(\chi^2) > 0.1$. 
At this point, the background with a photon escape was estimated to be
less than one event.

\begin{figure}
\centerline{\epsfig{file=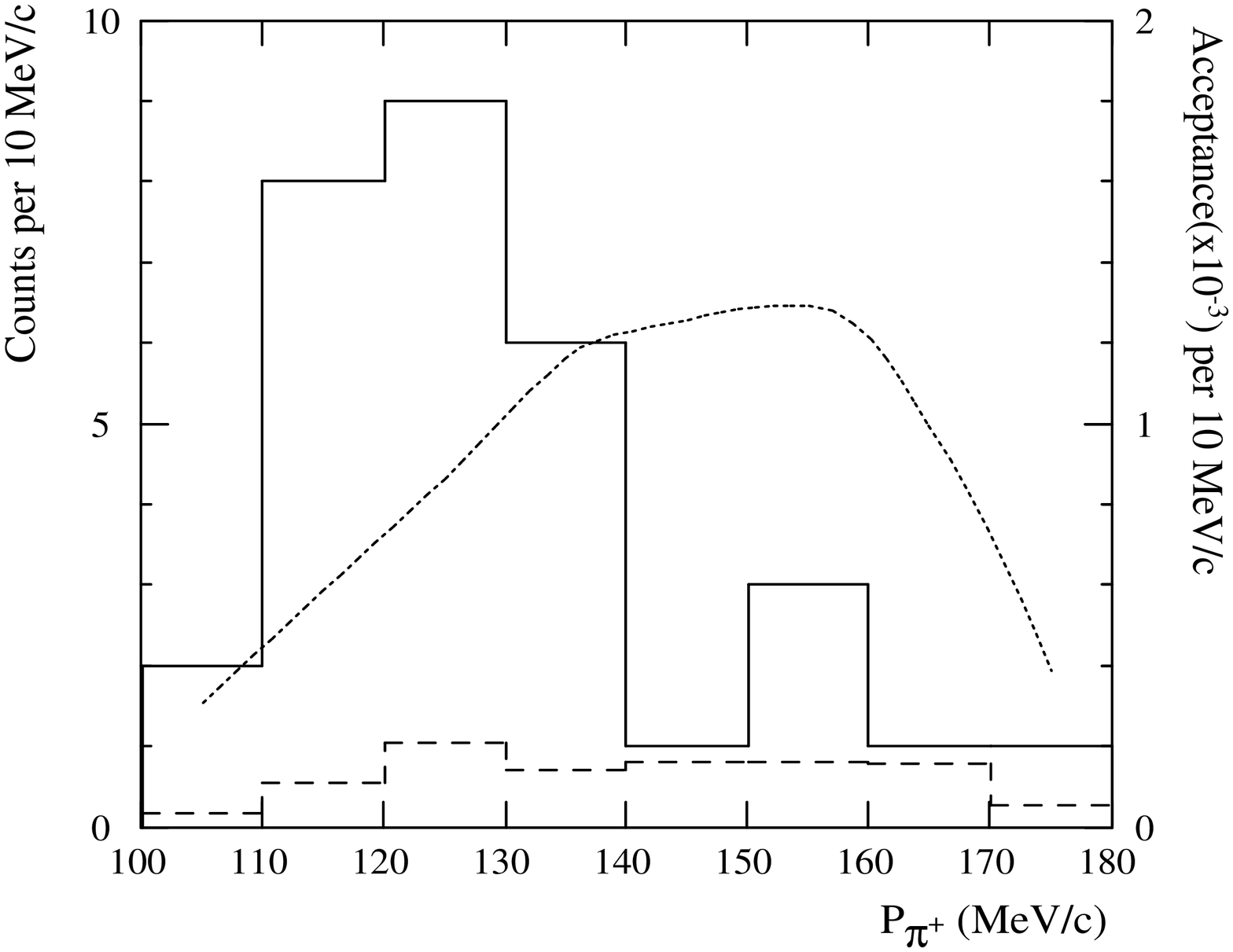,angle=-90,width=\textwidth}}
\caption{$\pi^+$ momentum distribution for 31
$K^+\rightarrow\pi^+\gamma\gamma$ candidates (solid) and for estimated
background events (dashed).  The acceptance is given by the dotted
line.}
\label{pggevent}
\end{figure}

Remaining background events were primarily due to $K_{\pi 3}$ and
$K_{\pi 2\gamma}$ decays with multiple photons fused in the barrel
calorimeter to form a single cluster. Background with overlapping
clusters was greatly reduced by rejecting events with cluster shapes
inconsistent with that of a single photon.  A typical single photon
cluster contained 4 -- 5 hit modules in the barrel calorimeter.  The
maximum discrepancy among $z$ measurements and the standard deviation
of $\phi$ measurements were calculated for each photon cluster, and
events were rejected if one of the cut variables exceeded the 72\%
acceptance point. The acceptance was measured using a clean 
single-photon cluster based on $K_{\pi 2}$
data with the same number of hit elements and a similar energy.  A
total of 31 $K^+\rightarrow\pi^+\gamma\gamma$ decay candidates
survived all the cuts (Fig.~\ref{pggevent}).  The residual background
was estimated by measuring the rejections of the $z$ and $\phi$
cuts. Each cut was inverted to enhance the
overlapping-cluster background, and the rejection of the other cut was
evaluated by assuming the two cuts were independent. The independence
of the cuts and the magnitude of the background level were confirmed
by a large statistics Monte Carlo simulation. The estimated number of
the total background was 5.1$\pm$3.3.

The acceptance was estimated by using Monte Carlo simulation and
calibration data sets.  The net acceptance for the $\pi\gamma\gamma1$
trigger and analysis was $(1.5 \pm 0.2) \times 10^{-4}$ which included
a factor 0.12 for the $\pi^+$ momentum being between 215 MeV/$c$ and
the end point 227 MeV/$c$ assuming the phase-space distribution. Under
this assumption, the 90\% C.L. upper limit was
B($K^+\rightarrow\pi^+\gamma\gamma$, phase-space distribution) $< 5.0
\times 10^{-7}$. 

The momentum dependence of the $\pi\gamma\gamma2$
acceptance is shown in Fig.~\ref{pggevent} together with the final
$\pi\gamma\gamma$ candidates and the estimated background spectrum. A
partial branching ratio for each 10-MeV/$c$ bin was calculated from
the background-subtracted signal divided by the number of the stopped
kaons times the acceptance for that bin. By summing these, a
model-independent branching ratio B($K^+\rightarrow\pi^+\gamma\gamma$,
100 MeV/$c$ $ < P_{\pi^+} < $ 180 MeV/$c$) was obtained to be $(6.0
\pm 1.5 \pm 0.7)
\times 10^{-7}$ where the first uncertainty is statistical and the
second is the estimated systematic uncertainty in the background and
the acceptance measurements based on a $K_{\pi 2}$ branching ratio
measurement.

\begin{figure}
\centerline{\epsfig{file=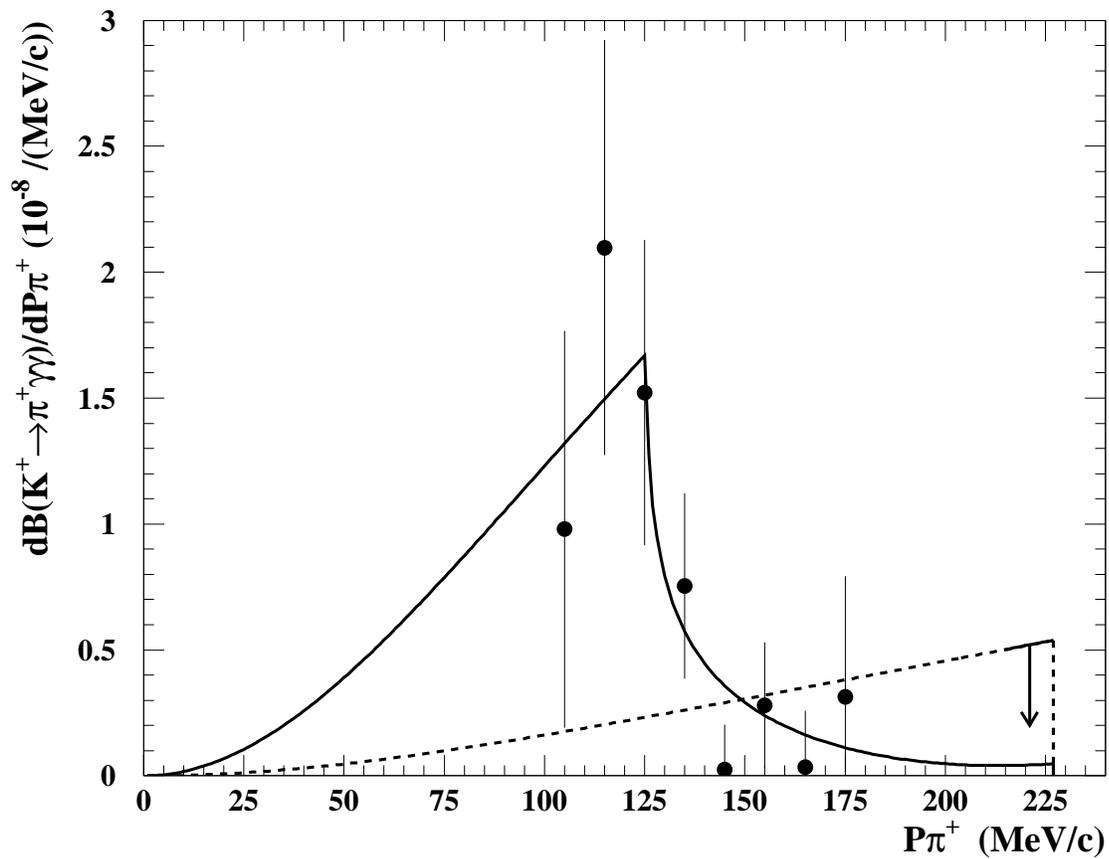,width=\textwidth}}
\caption{Measured $\pi^+$ momentum distribution for
$K^+\rightarrow\pi^+\gamma\gamma$ and the best fit to the data
(solid). The dashed line shows a phase-space distribution normalized to a
90\% C.L. upper limit obtained by the $\pi\gamma\gamma1$ analysis in the
region of 215 MeV/$c$ $ < P_{\pi^+} < 227$ MeV/$c$
indicated by the arrow.}
\label{pggspectrum}
\end{figure}

The measured $\pi^+$ spectrum was compared with ChPT predictions
integrated over 10-MeV/$c$ bins. A maximum likelihood fit of $\hat{c}$
to the spectrum using the absolutely-normalized rate was carried out 
to determine the
value of $\hat{c}$.  Without the unitarity corrections, the result was
$\hat{c} = 1.6 \pm 0.6$ with $\chi^2_{min.} = 6.3$ $(n.d.f. = 7)$.
With the corrections, the best fit was obtained for $\hat{c} = 1.8 \pm
0.6$, and the $\chi^2_{min.}$ improved to 4.6.  Thus, the data
support the inclusion of the unitarity corrections.  If we fit the
spectrum shape alone by introducing one free normalization factor, the
fits became equally good, with $\chi^2_{min.}/n.d.f. \sim 0.7$. The
corresponding $\hat{c}$ was $-0.6^{+ 1.4}_{- 1.1}$ and $0.7^{+ 1.7}_{-
1.1}$ for without and with the unitarity corrections, respectively.
Fig.~\ref{pggspectrum} shows the measured momentum spectrum together
with the spectrum shape given by the best fit ($i.e.$ $\hat{c}$ = 1.8
with the unitarity corrections). By assuming this spectrum shape, a
total branching ratio was estimated to be $(1.1 \pm 0.3 \pm 0.1)
\times 10^{-6}$ with the net acceptance in the $\pi\gamma\gamma2$
region being $3.8 \times 10^{-4}$.  The total branching ratio estimated
with a spectrum shape given by the other fits was within 11 \% of the
above value.

One of the consequences of the unitarity corrections is a non-zero
amplitude at the end point $P_{\pi^+} = 227$ MeV/$c$
($m_{\gamma\gamma} = 0$ MeV/$c^2$).  However, the predicted decay rate
at the end point is 8 times smaller than our 90\% C.L. upper limit in
this region, B($K^+\rightarrow\pi^+\gamma\gamma$, $P_{\pi^+} > 215$
MeV/$c$ ) = $6.1 \times 10^{-8}$, as shown in
Fig.~\ref{pggspectrum}. Therefore, the ChPT prediction with the unitarity
corrections is consistent with our $\pi\gamma\gamma1$ result.

\begin{figure}
\centerline{\epsfig{file=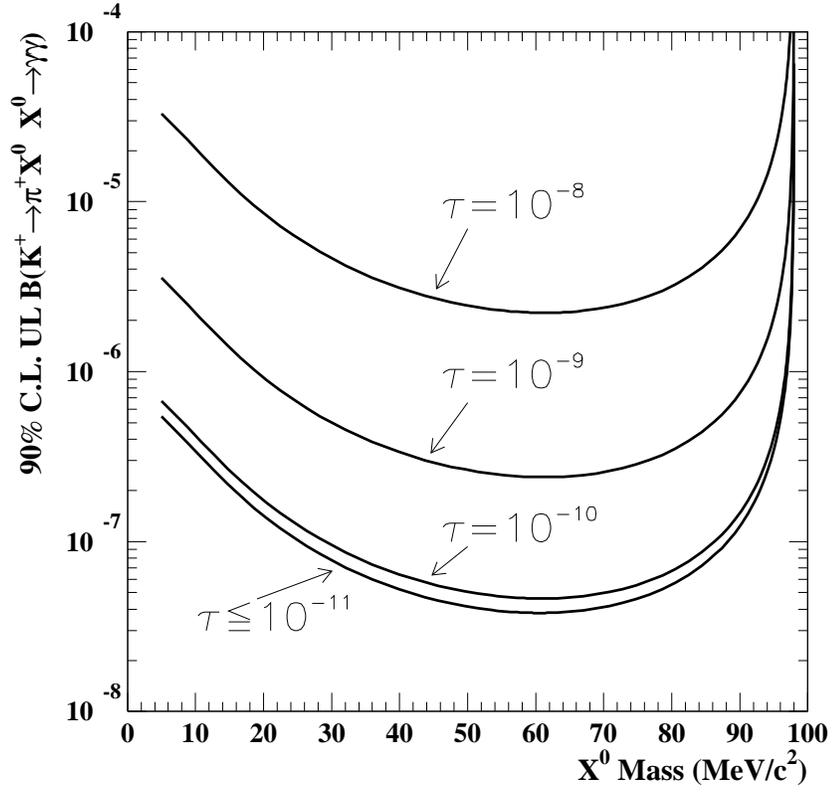,width=.75\textwidth}}
\caption{The 90\% C.L. upper limits for the branching ratio of $K^+
\rightarrow \pi^+ X^0,X^0 \rightarrow \gamma\gamma$ for different
$X^0$ lifetimes ($\tau_{X^0}$) as a function of mass ($m_{X^0}$).}
\label{uplimit}
\end{figure}

The $\pi\gamma\gamma1$ result also sets a 90 $\%$ confidence upper
limit on B($K^+ \rightarrow \pi^+ X^0,X^0 \rightarrow \gamma\gamma$),
where $X^0$ is any short-lived neutral particle with a mass smaller
than 100 MeV/$c^2$ decaying into two photons. Fig.~\ref{uplimit} shows
the limit as a function of $M_{X^0}$ for different lifetimes.


\acknowledgments

We gratefully acknowledge the dedicated efforts of the technical
staff supporting this experiment and of the Brookhaven AGS
Department.  This research was supported in part by the
U.S. Department of Energy under contracts DE-FG02-91ER40671,
W-7405-ENG-36, and grant DE-FG02-91ER40671, and by the Natural Sciences and
Engineering Research Council, by the National Research Council of
Canada, and by the Ministry of Education, Science, Sports and Culture
of Japan.



\newpage

\newpage

\newpage

\end{document}